\documentstyle[12pt]{article}
\parskip=\bigskipamount
\def\Alf{Alfv\'{e}n\ }
\def\Schr{Schr\"{o}dinger }
\def\DMVa{Deconinck, Meuris and Verheest 1993a}
\def\DMVb{Deconinck, Meuris and Verheest 1993b}
\def\DMVab{Deconinck, Meuris and Verheest 1993a, 1993b}
\def\pdv#1#2{\frac{\partial{#1}}{\partial{#2}}}
\def\beq{\begin{equation}}
\def\eeq{\end{equation}}
\def\beqa{\begin{eqnarray}}
\def\eeqa{\end{eqnarray}}

\def\vetga{\mbox{\boldmath $\gamma$}}
\def\vetze{\mbox{\boldmath $\zeta$}}
\def\vetBnul{{\bf B}_{\perp 0}}
\def\vetBpp{{\bf B}_\perp}

\def\vec#1#2{\left(\begin{array}{c}{#1}\\{#2}\end{array}\right)}
\def\intii#1#2{\int^{+\infty}_{-\infty}{#1}\ {#2}}
\def\vbvm#1#2#3#4{\left(\begin{array}{cc} {#1} & {#2} \\ {#3} &
{#4}\end{array}\right)}

\title{Complete integrability of a modified vector derivative nonlinear \Schr
equation}
\author{Ralph Willox$^1$, Willy Hereman$^2$ and Frank Verheest$^3$\\[5mm]
      $^1$Dienst Theoretische Natuurkunde, Vrije Universiteit Brussel,\\
          Pleinlaan 2, B--1050 Brussel, Belgium\\[5mm]
      $^2$Department of Mathematical and Computer Sciences,\\
          Colorado School of Mines,\\
          Golden, Colorado 80401--1887, USA\\[5mm]
      $^3$Sterrenkundig Observatorium, Universiteit Gent,\\
          Krijgslaan 281, B--9000 Gent, Belgium}

\begin{document}
\maketitle
\begin{abstract}
\noindent Oblique propagation of magnetohydrodynamic waves in warm plasmas is
described by a modified vector derivative nonlinear \Schr equation, if charge
separation in Poisson's equation and the displacement current in Amp\`ere's law
are properly taken into account.  This modified equation cannot be reduced to
the standard derivative nonlinear \Schr equation and hence its possible
integrability and related properties need to be established afresh.  Indeed,
the new equation is shown to be integrable by the existence of a
bi--Hamiltonian structure, which yields the recursion operator needed to
generate an infinite sequence of conserved densities.  Some of these have been
found explicitly by symbolic computations based on the symmetry properties of
the new equation.  Since the new equation includes as a special case the
derivative nonlinear \Schr equation, the recursion operator for the latter one
is now readily available.
\end{abstract}
\newpage

\section{Introduction}
The derivative nonlinear \Schr equation (DNLS) was first given by Rogister
(1971) for the nonlinear evolution of parallel \Alf waves in plasmas, and later
encountered in many different contexts by other authors, emerging as one of the
canonical nonlinear equations in physics.  The DNLS could also account for
slightly oblique propagation of \Alf waves (see {\it e.g.} Hada, Kennel and
Buti 1989), albeit at the price of neglecting two effects which might be
important in strongly magnetized astrophysical plasmas.  One is the deviation
from charge neutrality between the different plasma species, the other is the
influence of the displacement current in Amp\`ere's law.  Retaining these
effects results in a nonlinear vector evolution equation which differs from the
standard vector form of the DNLS by an extra linear term, and therefore was
called the modified vector derivative nonlinear \Schr equation (MVDNLS)
(\DMVab).

Recently, there has been a renewed interest in the use of the DNLS for certain
astrophysical plasmas (Spangler 1992, Spangler and Plapp 1992), assuming that
the case of slightly oblique propagation could easily be reduced to that of
parallel propagation.  To do so, one modifies the dependent variable, in this
case the perpendicular magnetic field, by including the static part as well.
Since the MVDNLS cannot be transformed into the DNLS itself, such an easy
transition from parallel to oblique propagation is not possible.  The reverse
is true, of course, the MVDNLS includes the DNLS as a special case, when we go
from oblique to parallel propagation, which amounts to dropping the bothersome
extra term.

That term, which distinguishes the MVDNLS from the DNLS, has implications for
the discussion of integrability and the possibility of deriving solitary wave
solutions for the MVDNLS.  The DNLS is well known to be completely integrable
(Kaup and Newell 1978), whereas for the MVDNLS we could only get certain
indications about its integrability (\DMVb).  The applicability of the
prolongation method (Kaup 1980), adapted to a vector nonlinear equation, and
the existence of some invariants (\DMVb) were indicative of complete
integrability, without giving a watertight proof.

In the present paper we show that the MVDNLS possesses a bi--Hamiltonian
structure, and hence through the resulting recursion operator an infinite
sequence of conserved densities.  Interestingly enough, to the best of our
knowledge there seems to be no proof in the literature that the DNLS itself has
a bi--Hamiltonian structure, although this expected property is mentioned
sometimes without further details nor references (Oevel and Fuchssteiner 1992).
The existence of an infinite sequence of conserved densities is proved by Kaup
and Newell (1978) for the DNLS, without showing the bi--Hamiltonian character.
Our constructions, formulas and conclusions for the MVDNLS immediately apply
to the DNLS.  Hence, without extra work, the explicit forms of both
Hamiltonians and the recursion operator for the DNLS are now available.

In \S 2 we recall the form of the MVDNLS and list in \S 3 some of the conserved
densities, which were found in an ad hoc fashion with the help of a symbolic
program and by looking at the symmetry properties of the equation.  The
knowledge of these conserved densities turned out to be beneficial for an easy
construction of the appropriate Hamiltonians in \S 4.  In \S 5 the
bi--Hamiltonian structure and the recursion operator are derived and with it
we established the existence of an infinite sequence of conserved densities,
needed to guarantee complete integrability.  \S 6 is devoted to a short
discussion of the implications for the usual DNLS, and in \S 7 we draw some
conclusions.

\section{MVDNLS}
The MVDNLS is, after the necessary scaling and Galilean transforms to cast it
in its simplest dimensionless form, given by
\beq 
\pdv{\vetBpp}{t} + \pdv{}{x} (B_\perp^2 \vetBpp) + \alpha \vetBnul \vetBnul
\cdot \pdv{\vetBpp}{x} + {\bf e}_x \times \pdv{^2 \vetBpp}{x^2} = {\bf 0},
\end{equation}
where the parameter $\alpha$ characterizes the extra term which distinguishes
the MVDNLS from the DNLS.  In (1) $\vetBpp$ stands for the perpendicular
magnetic field, which includes both the wave contributions and the static
perpendicular field due to the oblique propagation with respect to the total
external magnetic field.  The direction of wave propagation is along the
$x$--axis.  If the third term is absent, we can project (1) onto axes
perpendicular to the direction of wave propagation, introduce a new complex
variable from the components of $\vetBpp$,
\beq 
\phi_\pm = B_y \pm i B_z,
\eeq
and combine the projections to obtain the DNLS in standard scalar form,
\beq 
\pdv{\phi_\pm}{t} + \pdv{}{x} (|\phi_\pm|^2 \phi_\pm) \pm i \pdv{^2
\phi_\pm}{x^2} = 0.
\end{equation}
The $\pm$ signs in (2) and (3) are correlated.  The DNLS can account for
oblique propagation, provided $\alpha$ is zero, {\it i.e.\/} imposing charge
neutrality to all orders and neglecting the displacement current in Amp\`ere's
law (\DMVa).  The bothersome third term in (1) also disappears if $\vetBnul$
is zero, in the case of strictly parallel propagation.

At this stage, it is worth recalling that the DNLS has constant-amplitude
solutions of the form
\begin{eqnarray} 
B_y & = & a \cos(k x - \omega t), \nonumber \\[-4mm]\\[-3mm]
B_z & = & \mbox{} \pm a \sin(k x - \omega t), \nonumber
\end{eqnarray}
with $a$ an arbitrary constant.  As can easily be checked, there are no
constant amplitude solutions to (1) besides the trivial case with $\vetBpp =
\vetBnul$, if $\vetBnul \neq 0$.  This means that any sort of separation into
left and right circularly polarized waves as for the DNLS is doomed to fail.
Of course, for oblique propagation with $\alpha = 0$, the circular polarization
given by the DNLS is only apparent, since $\vetBpp$ includes the static part
$\vetBnul$, and this shift leads in reality to elliptical polarization for the
perpendicular wave field (Spangler and Plapp 1992).  In addition, the MVDNLS
has a class of stationary solitary wave solutions which the DNLS does not have,
the subalfv\'enic modes, which have totally different properties compared to
the known stationary solutions of the DNLS (\DMVb).

We know that the DNLS is completely integrable in the sense that it is
possesses an infinite series of conserved densities, which can be constructed
explicitly.  As the MVDNLS reduces for $\alpha \rightarrow 0$ to the vector
form of the DNLS, there is hope to prove integrability for the MVDNLS.  There
are, however, two major complications with the MVDNLS: it has a vector
character which contrary to the DNLS cannot be transformed away, and the
boundary values at infinity are not zero (at least not in the physical model
for which the nonlinear evolution equation was derived!).

One encounters in the literature quite a variety of methods to investigate
symmetries, to construct conservation laws, or to establish integrability of
nonlinear equations via direct or inverse methods.  While in principle these
methods could be used, one rarely sees worked examples involving vector
equations.  Furthermore, the nonzero boundary conditions for the MVDNLS are an
additional hurdle.

\section{Integrability and conserved densities}
The first step in any treatment is to try to prove integrability, or at least
collect sufficiently compelling evidence.  For instance, one could check if the
equation passes the Painlev\'e test.  However, here again the nonscalar
character of the MVDNLS prevents straightforward application of this otherwise
so useful test (Fordy 1990).   A different way to ascertain integrability, at
least as convincing, is indicated by Kaup (1980) and Fordy (1990).  It involves
the application of the prolongation method due to Estabrook and Wahlquist.
This method was adapted successfully to the nonlinear equation at hand (\DMVb).

Yet another important tool to determine integrability of a nonlinear PDE is
finding a sufficiently large number of conservation laws of the form
\beq 
\pdv{\rho}{t} + \pdv{J}{x} = 0,
\eeq
where $\int_{-\infty}^{+\infty} \rho\, dx$ is the conserved quantity with
density $\rho$ and associated flux $J$.  We have assumed functions $u$ and $v$
which are fast decreasing at infinity to symmetric but in the case of $u$
non--zero boundary conditions.  As is well known from many other examples, the
first conservation law comes from rewriting the equations themselves in the
form of a conservation law.  For the MVDNLS this yields after projection
\beqa 
&& u_t + \left( u(u^2 + v^2) + \beta u - v_x \right)_x = 0, \nonumber
\\[-3mm]\\[-3mm]
&& v_t + \left( v(u^2 + v^2) + u_x \right)_x = 0, \nonumber
\eeqa
where $u$ and $v$ denote the components of $\vetBpp$ parallel and perpendicular
to $\vetBnul$, and $\beta=\alpha B^2_{\perp 0}$.  As usual, subscripts refer
to partial derivatives with respect to $t$ and $x$.  Whereas (6) amounts to a
vector conservation law, the other ones we have derived through constructive
procedures are all scalar ones, involving powers of $(u^2 + v^2)$, as presented
in (9)--(13).

To see how to proceed, we follow ideas exposed in more detail in Verheest and
Hereman (1994), and briefly discuss the scaling or symmetry properties of the
equations.  These can be used to obtain information about polynomial conserved
densities and the building blocks they are made off.  The scaling of (6) is
such that
\beq 
u \sim v, \qquad\qquad {\partial \over \partial t} \sim {\partial^2 \over
\partial x^2}, \qquad\qquad u^2 \sim v^2 \sim \beta \sim {\partial \over
\partial x} .
\eeq
We may restrict ourselves to building blocks which belong to the same class
under the mentioned scaling, since for any mixed conserved quantity which one
could derive, the freedom implied in the scaling would split that quantity in
several conserved quantities, each with building blocks of the same scaling.
Thus there is a straightforward and logical way to construct invariant
quantities (Miura, Gardner and Kruskal (1968)).  Moreover and without loss of
generality, we may remove any density (or part thereof) that is a total
$x$--derivative, for these are trivially conserved.  In addition, for $\beta=0$
(6) is invariant under the substitution
\beq 
u \rightarrow v, \qquad\qquad v \rightarrow {} - u.
\eeq
In every conserved density the part without factors $\beta$ will have to obey
this additional symmetry.

At the quadratic level in $u$ and $v$, the only possibility is $u^2 + v^2$ or
(9) given below, due to the rule in (8).  Cubic terms in $u$ and $v$ are not
possible at all.

Starting then with a candidate density containing the building block $(u^2 +
v^2)^2$, one has four factors $u$ or $v$, and one could add a combination of
the form $u^3 v - u v^3$.  Keeping the scaling (7) in mind, quartic terms are
also equivalent to two factors $u$ and/or $v$ and one derivation, or to two
factors $u$ and/or $v$ and one factor $\beta$.   Two factors $u$ and/or $v$
with one derivation can only lead to a non-trivial building block of the
structure $u v_x - v u_x$, if we keep all the preceding remarks in mind.  With
one factor $\beta$ we could in principle have a linear combination of $u^2$,
$u v$ and $v^2$.  This exhausts the building blocks at this order and it is
then for the MATHEMATICA program we wrote to determine the necessary
coefficients.  This leads to (10).  Obviously, we can go on like this for
higher orders.

Labelling the conserved densities $\rho_n$, with $n$ the corresponding power
of $(u^2 + v^2)$, we obtained
\beqa 
\rho_1 & = & u^2 + v^2, \\[3mm]
\rho_2 & = & \frac{1}{2} (u^2+v^2)^2 - uv_x + u_xv + \beta u^2, \\[3mm]
\rho_3 & = & \frac{1}{4} (u^2+v^2)^3 + \frac{1}{2}(u_x^2 + v_x^2)
- u^3v_x + v^3u_x + \frac{\beta}{4}(u^4-v^4), \\[3mm]
\rho_4 & = & \frac{1}{4}(u^2+v^2)^4 - \frac{2}{5} (u_xv_{xx}-u_{xx}v_x)
+ \frac{4}{5} (uu_x+vv_x)^2  \nonumber \\
& & \mbox{} + \frac{6}{5} (u^2+v^2)(u_x^2+v_x^2) - (u^2+v^2)^2(uv_x-u_xv) \\
& & \mbox{} + \frac{\beta}{5} (2u_x^2 - 4u^3v_x + 2u^6 + 3u^4v^2 - v^6)
+ \frac{\beta^2}{5} u^4, \nonumber \\[3mm]
\rho_5 & = & \frac{7}{16} (u^2+v^2)^5 \! +\! \frac{1}{2}(u_{xx}^2+v_{xx}^2)
\!-\! \frac{5}{2}(u^2+v^2)(u_xv_{xx}
\!-\! u_{xx}v_x) + 5(u^2+v^2)(uu_x+vv_x)^2 \nonumber \\
& & \mbox{} + \frac{15}{4} (u^2+v^2)^2(u_x^2+v_x^2)^2 - \frac{35}{16}
(u^2+v^2)^3(uv_x-u_xv)  \nonumber \\[-4mm]\\
& & \mbox{} + \frac{\beta}{8} ( 5u^8 + 10u^6v^2 -
10u^2v^6 - 5v^8 + 20u^2u_x^2 - 12u^5v_x + 60uv^4v_x - 20v^2v_x^2 ) \nonumber
\\
& & \mbox{} + \frac{\beta^2}{4} (u^6+v^6). \nonumber
\eeqa
Note that we have not yet included the renormalization constants needed to
ensure the boundedness of the conserved quantities obtained from the above
listed densities.  We will come back to this point in the following paragraph.

\section{Hamiltonian structure}
Let us start by pointing out some of the principal ingredients of the
Hamiltonian structure of evolution equations, adapted where necessary to vector
quantities.

The system (6) is said to possess a Hamiltonian structure (Olver 1980), if
there exists a so called Hamiltonian operator $\Theta$ (Fokas 1987) (sometimes
called implectic operator, see {\it e.g.\ }Fuchssteiner and Fokas 1981) and a
(2-component) gradient vector function $\vetga_H(u,v)$ such that (6) can be
written in the form:
\beq 
\vec{u}{v}_t = \Theta \cdot \vetga_H\ .
\eeq
The operator $\Theta$ is a Hamiltonian operator if it is skew-symmetric
\beq 
\langle \Theta\cdot {\bf a} , {\bf b} \rangle = - \langle {\bf a} , \Theta\cdot
{\bf b} \rangle
\eeq
with respect to the scalar product
\beq 
\langle {\bf f} , {\bf g} \rangle = \intii{{\bf f}(x) \cdot {\bf g}(x)}{dx}\
,
\eeq
and if it satisfies a ``Jacobi-like'' identity.  The precise form (A.1) of that
identity is given in the Appendix, since it is not of immediate importance
here.  A vector function $\vetga_H(u,v)$ is a gradient function if its
Fr\'echet derivative is symmetric with respect to the scalar product (16):
\beq 
\langle \vetga^{'}_H[{\bf a}] , {\bf b} \rangle = \langle {\bf a} ,
\vetga^{'}_H[{\bf b}] \rangle\ .
\eeq
Recall that the Fr\'echet derivative of a vector function
$\vetga_H(u,v)=\vec{\gamma_1(u,v)}{\gamma_2(u,v)}$ in a direction
$\vec{\xi}{\eta}$ is given by
\beq 
\vetga^{'}_H\left[\vec{\xi}{\eta}\right] =
\pdv{}{\varepsilon}\left.\vec{\gamma_1(u+\varepsilon \xi,v) +
\gamma_1(u,v+\varepsilon \eta)}{\gamma_2(u+\varepsilon \xi,v) +
\gamma_2(u,v+\varepsilon \eta)}\right|_{\varepsilon=0}\ .
\eeq
The Hamiltonian, or Hamiltonian functional, $H = \int^{+\infty}_{-\infty}
\rho(u,v) dx$ giving rise to the gradient function $\vetga_H(u,v)$ through the
identity (Berger 1977)
\beq 
H^{'}[\vetze] = \intii{\rho^{'}[\vetze]}{dx} \equiv \langle \vetga_H, \vetze
\rangle\ ,
\eeq
can be recovered from this gradient in the following way:
\beq 
H = \int^1_0 \left[ \langle \vetga_H(\lambda u, \lambda v) , \vec{u}{v} \rangle
 - C \right]  d\lambda\ ,
\eeq
where the constant $C$ is chosen such that the integral giving $H$ is bounded.
Such a Hamiltonian formulation admits a Poisson bracket,
\beq 
\left\{A,B\right\} \equiv \langle \vetga_A, \Theta\cdot \vetga_B \rangle\ ,
\eeq
defining the time evolution of a functional $A$ of $u$ and $v$:
\beqa 
\frac{d A}{dt} = \pdv{A}{t} + A^{'}\left[\vec{u}{v}_t\right] & = & \pdv{A}{t}
+ \langle \vetga_A , \vec{u}{v}_t \rangle\nonumber \\[-4mm]\\[-3mm]
& = & \pdv{A}{t} + \langle \vetga_A, \Theta\cdot \vetga_H \rangle = \pdv{A}{t}
+ \left\{A,H\right\},\nonumber
\eeqa
using formulas (19) and (14).

Since $\Theta$ is skew-symmetric, and due to identity (A.1), the bracket (21)
possesses all the characteristics of a standard Poisson bracket, except for the
``Leibniz-like'' expulsion property which cannot be properly defined in the
case of functionals. Clearly, if autonomous, the Hamiltonian $H$ itself is a
conserved quantity for the evolution equation (14). Since (6) can be written
as a conservation law, an apparent Hamiltonian formulation is the following:
\beqa 
\vec{u}{v}_t & = & \Theta_2\cdot \vetga_2 ,\nonumber \\[-2mm]\\[-1mm]
\Theta_2 = - \vbvm{\partial_x}{0}{0}{\partial_x} & , & \vetga_2 =
\vec{(u^2+v^2) u - v_x + \beta u}{(u^2+v^2) v + u_x}\ ,\nonumber
\eeqa
where $\partial_x$ denotes the partial derivative with respect to $x$.

$\Theta_2$ is easily seen to be a Hamiltonian operator since it is a constant
({\it i.e.\/} not depending on $u$ or $v$), skew-symmetric operator (see
Appendix).

It is straightforward to verify that $\vetga_2$ satisfies (17) and thus leads
to a Hamiltonian functional $H_2$ given by formula (20):
\beqa 
H_2 & = & \int^1_0 d\lambda \intii{\left[\lambda^3 (u^2+v^2)^2 + \lambda (u_x
v -u v_x + \beta u^2) - C_2 \right]}{dx}\nonumber \\[-2mm]\\[-1mm]
& = & \intii{\left[\frac{(u^2+v^2)^2}{4} + \frac{u_x v - u v_x}{2} +
\frac{\beta}{2} u^2 - C_2 \right]}{dx}\ .\nonumber
\eeqa
Comparison with formula (10) shows that it is the conserved density $\rho_2$
which give rise to this Hamiltonian.  One may wonder if the conserved density
$\rho_1$ in (9) can be linked to a Hamiltonian functional as well. Let us
define $H_1$ by
\beq 
H_1 = \intii{\left[\frac{\rho_1}{2} - C_1\right]}{dx} =
\intii{\left[\frac{u^2+v^2}{2} - C_1\right]}{dx}\ .
\eeq
Using (19), the gradient $\vetga_1$ of this functional is found to be:
\beq 
\vetga_1 = \vec{u}{v}\ .
\eeq
The challenge is to find a Hamiltonian operator $\Theta_1$ such that (6) can
be recast into the form (14) using $\vetga_1$.  The MVDNLS equation would then
have a second Hamiltonian formulation and thus possess a bi--Hamiltonian
structure.

\section{Bi--Hamiltonian structure and recursion operator}
If an evolution equation admits a bi--Hamiltonian formulation (Magri 1978)
\beq 
\vec{u}{v}_t = \Theta_1 \cdot \vetga_1 = \Theta_2 \cdot \vetga_2\ ,
\eeq
and if it is possible to show that $\Theta_1$ and $\Theta_2$ are compatible
Hamiltonian operators ({\it i.e.\/} $\Theta_1 + \Theta_2$ is again a
Hamiltonian operator), and if one of the operators, say $\Theta_2$ is
invertible, then the operator
\beq 
R = \Theta_1 \cdot \Theta_2^{-1}
\eeq
is a hereditary recursion operator (Fuchssteiner 1979, Fuchssteiner and Fokas
1981) for that evolution equation.

The formal adjoint of $R$ with respect to the scalar product (16)
\beq 
R^{\dagger} = \Theta_2^{-1} \cdot \Theta_1
\eeq
then maps gradients of conserved quantities into gradients, provided the
operator $R$ is injective (Fuchssteiner and Fokas 1981, and also Appendix),
thus defining an infinite sequence of conserved quantities, all in involution
with respect to the Poisson bracket (21).

In our case we have already found one Hamiltonian structure with an invertible
Hamiltonian operator $\Theta_2$:
\beq 
\Theta_2^{-1} = - \vbvm{\partial_x^{-1}}{0}{0}{\partial_x^{-1}}\ ,
\eeq
where $\partial_x^{-1}$ denotes the inverse of the $\partial_x$ operator, such
that $\partial_x \partial_x^{-1} = \partial_x^{-1} \partial_x = 1$.

A first step towards finding a second Hamiltonian structure is the construction
of a skew-symmetric operator $\Theta_1$ which casts (6) into Hamiltonian form
with the gradient $\vetga_1$ in (26). The most general parametrization
(involving a single $\partial_x^{-1}$ operator) which satisfies the above
constraints is
\beq 
\Theta_1 = - \vbvm{\theta_{11}}{\theta_{12}}{\theta_{21}}{\theta_{22}}\ ,
\eeq
with
\beqa 
\theta_{11} & = & \beta \partial_x + c_1 v \partial_x v + c_3 u \partial_x u
+ (6-4 c_3) u_x \partial_x^{-1} u_x \ ,\nonumber\\
\theta_{12} & = & -\partial_x^2 + (\frac{3}{2} - \frac{c_1+c_2}{2} -
\frac{c_5}{4}) u v \partial_x + (\frac{1}{2} - \frac{c_1-c_2}{2}
+\frac{c_5}{4}) u v_x \nonumber \\
& & \hskip 10mm + (1 - c_1 -\frac{c_5}{2}) u_x v + c_5 u_x \partial_x^{-1} v_x
\ , \nonumber  \\[-4mm]\\[-3mm]
\theta_{21} & = & \partial_x^2 + (\frac{3}{2} - \frac{c_1+c_2}{2} -
\frac{c_5}{4}) u v \partial_x + (\frac{1}{2} + \frac{c_1-c_2}{2}
+\frac{c_5}{4}) u_x v \nonumber\\
& & \hskip 10mm + (1 - c_1 -\frac{c_5}{2}) u v_x + c_5 v_x \partial_x^{-1} u_x\
,\nonumber\\
\theta_{22} & = & c_2 u \partial_x u + c_4 v \partial_x v + (6-4
c_4) v_x \partial_x^{-1} v_x\ .\nonumber
\eeqa
At this point one could of course try to find values of the parameters for
which this operator is actually a Hamiltonian operator.  However, this proves
to be a formidable task (see Appendix). A better approach is to concentrate on
the action of $R^{\dagger}$ on the gradient functions obtained so far. Because
of the bi--Hamiltonian structure (27) of the equations, $R^{\dagger}$ maps
$\vetga_1$ into $\vetga_2$:
\beq 
\vetga_2 = \Theta_2^{-1} \cdot \Theta_1 \cdot \vetga_1 = R^{\dagger} \cdot
\vetga_1\ .
\eeq
Suppose $R^{\dagger}$ actually is the formal adjoint of a recursion operator
for the MVDNLS equation, then it will map $\vetga_2$ into yet another gradient
function. If the conserved quantity corresponding to this gradient has to be
polynomial in $u , v$ and their derivatives, then the values of the parameters
in (32) have to be such that $\Theta_1 \cdot \vetga_2$ is a total
$x$-derivative:
\beq 
\vetga_3 = \Theta_2^{-1} \cdot \Theta_1 \cdot \vetga_2 = -
\vbvm{\partial_x^{-1}}{0}{0}{\partial_x^{-1}} \cdot \Theta_1 \cdot \vetga_2\
{}.
\eeq
This requirement uniquely determines all five parameters in (32), namely $c_1
= c_2 = 0, c_3 = c_4 = - c_5 = 2$, thus giving us a single candidate for a
second Hamiltonian operator:
\beq 
\Theta_1 \equiv - \vbvm{\beta \partial_x + 2 u \partial_x u - 2 u_x
\partial_x^{-1} u_x}{-\partial_x^2 + 2 v \partial_x u - 2 u_x \partial_x^{-1}
v_x}{\partial_x^2 + 2 u \partial_x v - 2 v_x \partial_x^{-1} u_x}{2 v
\partial_x v - 2 v_x \partial_x^{-1} v_x}\ ,
\eeq
together with the gradient
\beq 
\vetga_3 = \beta \vetga_2 + \vec{\beta u^3 - u_{xx} + \frac{3}{2} u^5 +
\frac{3}{2} u v^4 + 3 u^3 v^2 - 3 v^2 v_x - 3 u^2 v_x}{-\beta v^3 - v_{xx} +
\frac{3}{2} v^5 + \frac{3}{2} u^4 v + 3 u^2 v^3 + 3 u_x v^2 + 3 u^2 u_x}\ .
\eeq
Redefining $\vetga_3$ as $\vetga_3 - \beta \vetga_2$, it is straightforward to
show that it satisfies (17) and corresponds to the Hamiltonian functional
\beq 
H_3 = \intii{\left[\frac{(u^2+v^2)^3}{4} + u_x v^3 - u^3 v_x - \frac{1}{2} (u
u_{xx} + v v_{xx}) + \frac{\beta}{4} (u^4 - v^4) - C_3 \right]}{dx}\ .
\eeq
The density function associated with this functional is, up to partial
integration, the conserved density $\rho_3$ given in (11). Hence, it follows
that $R^{\dagger}$ maps $\vetga_2$ into the gradient of a conserved quantity
({\it i.e.\/}, $H_3 + \beta H_2$), suggesting that $R$ indeed is a hereditary
recursion operator for the MVDNLS equation.

In the Appendix it is proven that $\Theta_1$ is a Hamiltonian operator and that
it is compatible with $\Theta_2$.  Thus, we have shown that
\beq 
R = \vbvm{\beta + 2 u^2 + 2 u_x \partial_x^{-1} u}{-\partial_x + 2 u v + 2 u_x
\partial_x^{-1} v}{ \partial_x + 2 u v + 2 v_x \partial_x^{-1} u}{2 v^2 + 2 v_x
\partial_x^{-1} v}
\eeq
which follows from (28), (30) and (35), is a hereditary recursion operator for
the MVDNLS equation:
\beq 
\vec{u}{v}_t = \Theta_1 \cdot \Theta_2^{-1} \cdot \vec{u}{v}_x = R \cdot
\vec{u}{v}_x\ .
\eeq
The recursion operator therefore defines a hierarchy of integrable evolution
equations
\beq 
\vec{u}{v}_t = R^n \cdot \vec{u}{v}_x\ ,
\eeq
all sharing an infinite sequence of conserved quantities, the gradients of
which are generated by the formal adjoint of $R$:
\beq 
R^{\dagger} = \vbvm{\beta + 2 u \partial_x^{-1} u \partial_x}{-\partial + 2 u
\partial_x^{-1} v \partial_x}{\partial_x + 2 v \partial_x^{-1} u \partial_x}{2
v \partial_x^{-1} v \partial_x}\ .
\eeq

\section{The DNLS as special case}
For $\beta = 0$ we recover from the MVDNLS the usual DNLS itself.  Hence
$R|_{\beta = 0}$ or
\beq 
R = \vbvm{2 u^2 + 2 u_x \partial_x^{-1} u}{-\partial_x + 2 u v + 2 u_x
\partial_x^{-1} v}{ \partial_x + 2 u v + 2 v_x \partial_x^{-1} u}{2 v^2 + 2 v_x
\partial_x^{-1} v}
\eeq
is the recursion operator for DNLS, since it can be seen that the $\beta$ term
in $\Theta_1$ does not alter the fact that it is a Hamiltonian operator (see
Appendix). The Hamiltonian formulation (23) reduces to the one given by Kaup
and Newell (1978) when $\beta = 0$. However, to the best of our knowledge, the
second Hamiltonian formulation is a new result, not found in the literature,
and neither is the recursion operator (42).

\section{Conclusions}
Slightly oblique propagation of \Alf waves in strongly magnetized plasmas is
described by a nonlinear vector evolution equation which differs from the
vector form of the DNLS by an extra linear term, if one retains both the
deviation from charge neutrality between the different plasma species and the
influence of the displacement current in Amp\`ere's law.  The resulting
modified vector derivative nonlinear \Schr equation cannot be transformed into
the DNLS itself, and this has implications for the discussion of integrability
and the possibility of finding conserved densities and solitary wave solutions.
Of course, the reverse is true: the MVDNLS includes the DNLS as a special case.

While in a previous paper the applicability of the prolongation method, adapted
to a vector nonlinear equation, and the existence of some invariants were
indicative of complete integrability, in the present paper we have shown that
the MVDNLS indeed possesses a bi--Hamiltonian structure, and hence, through the
resulting recursion operator, an infinite sequence of conserved densities.  We
were guided in this by the explicit symbolic computation of the first seven
conserved densities.

Surprisingly enough, there seems to be no proof in the literature that the DNLS
itself has a bi--Hamiltonian structure, although the existence of an infinite
sequence of conserved densities was known.  As the DNLS emerges as a special
case of the MVDNLS studied here, we now also have the explicit form of the
recursion operator for the DNLS, besides the proof of its bi--Hamiltonian
structure.

\section*{Acknowledgements}  The National Fund for Scientific Research
(Belgium) is thanked for a special research grant, which made the stay of WH
at the Universiteit Gent possible, during which this work was initiated.  One
of the authors (RW) is a Senior Research Assistant at the National Fund for
Scientific Research (Belgium).  WH acknowledges support of the National Science
Foundation of America under Grant CCR--9300978.
\newpage
\setcounter{equation}{0}
\renewcommand{\theequation}{A.\arabic{equation}}
\section*{Appendix}

A skew-symmetric operator $\Theta$ is a Hamiltonian operator if and only if it
satisfies the identity (Fuchssteiner and Fokas 1981)
\beq 
\langle {\bf a} , \Theta^{'}[\Theta \cdot {\bf c}] \cdot {\bf b}\rangle +
\langle {\bf b} , \Theta^{'}[\Theta \cdot {\bf a}] \cdot {\bf c}\rangle +
\langle {\bf c} , \Theta^{'}[\Theta \cdot {\bf b}] \cdot {\bf a}\rangle \equiv
0\ ,
\eeq
where the scalar product was defined in (16).  Since this identity involves
taking the Fr\'echet derivative of $\Theta$ in a certain direction, it is
trivially satisfied if $\Theta$ does not depend upon $u$ or $v$, {\it i.e.\/}
if it is a ``constant'' operator.  Hence, every constant skew-symmetric
operator is a Hamiltonian operator.

Verifying that $\Theta_1$ satisfies (A.1) is a very cumbersome task.  A method
which makes such a verification a lot more tractable relies on defining a
functional multivector (see Olver 1986)
\beq 
{\cal Z} = \frac{1}{2} \int \{
\left(\begin{array}{cc}{\xi\ }{\eta}\end{array}\right) \wedge \Theta \cdot
\vec{\xi}{\eta}\}\ dx\ .
\eeq
Using this formalism, it can be shown (Olver 1986) that the identity (A.1) can
be recast into the form
\beq 
{\cal Z}^{'}\left[ \Theta \cdot \vec{\xi}{\eta} \right] \equiv 0\ ,
\eeq
assuming, by definition, that $\xi$ and $\eta$ are independent of $u , v$ or
their derivatives.

In our case we define the multivector
\beqa 
{\cal Z}_1 & = & \frac{1}{2} \int \{ (\beta + 2 u^2) \xi \wedge \xi_x - 2 u_x
\xi \wedge \partial_x^{-1} u_x \xi - 2 \xi \wedge \eta_{xx} + 4 u v \xi \wedge
\eta_x \nonumber \\[-4mm]\\[-3mm]
&& \phantom{\frac{1}{2} \int} {} + 4 u_x v \xi \wedge \eta -4 u_x \xi \wedge
\partial_x^{-1} v_x \eta + 2 v^2 \eta \wedge \eta_x - 2 v_x \eta \wedge
\partial_x^{-1} v_x \eta \}\ dx\ .\nonumber
\eeqa
It is a straightforward, albeit tedious, calculation to show that
\beq 
{\cal Z}_1^{'}\left[\Theta_1\cdot\vec{\xi}{\eta}\right] = 0\ ,
\eeq
thus proving that $\Theta_1$ is a Hamiltonian operator.

In the same manner it can be shown that
\beq 
{\cal Z}_1^{'}\left[\Theta_2\cdot\vec{\xi}{\eta}\right] = 0\ ,
\eeq
and hence that $\Theta_1 + \Theta_2$ is a Hamiltonian operator as well, {\it
i.e.\/} thus $\Theta_1$ and $\Theta_2$ are compatible Hamiltonian operators.
Note that since the only term in ${\cal Z}_1$ depending on $\beta$ is
independent of $u$ or $v$, its presence (or absence) does not alter the
properties (A.5) or (A.6). Consequently, the above considerations also hold in
the special case of the standard DNLS equation.

Concerning the injective nature of the recursion operator $R$, which is needed
to guarantee that $R^{\dagger}$ maps gradients to gradients (see Proposition
2 in Fuchssteiner and Fokas 1981), it suffices to consider the action of $R$
on the space ${\cal S}_P$ of (2-component) vector functions with polynomial
components in $u,v$ and their derivatives, which are all ${\cal C}^{\infty}$
functions fast decreasing at infinity to symmetric but in the case of $u$
non--zero boundary conditions.  It is easily seen that the kernel of $R$ in
this space consists of the elements which are of zero polynomial degree. Since
$R$ either raises the polynomial degree of the different terms by two or leaves
it invariant, $R$ can be made injective by restricting the domain to the space
${\cal S}_{P_0} \subset {\cal S}_P$ without zero degree elements.

\newpage
\section*{References}
\begin{description}
\item  Berger M S 1977, {\it Nonlinearity and Functional Analysis} (Academic
Press, N Y)
\item  Deconinck B, Meuris P and Verheest F 1993a, {\it J. Plasma Phys.} {\bf
50}, 445--455
\item  Deconinck B, Meuris P and Verheest F 1993b, {\it J. Plasma Phys.} {\bf
50}, 457--476
\item  Fokas A S 1987, {\it Stud.\ in Appl.\ Math.} {\bf 77}, 253--299
\item  Fordy A 1990, in: {\it Soliton theory: a survey of results} (edited by
A Fordy, Manchester University Press, Manchester) 403--426
\item  Fuchssteiner B 1979, {\it Nonlin.\ Anal., Theor.\ Meth.\ \& Appl.} {\bf
3}, 849--862
\item  Fuchssteiner B and Fokas A S 1981, {\it Physica D} {\bf 4}, 47--66
\item  Hada T, Kennel C F and Buti B 1989, {\it J. Geophys.\ Res.} {\bf 94},
65--77
\item  Kaup D J 1980, {\it Physica D} {\bf 1}, 391--411
\item  Kaup D J and Newell A C 1978, {\it J. Math.\ Phys.} {\bf 19}, 798--801
\item  Magri F 1978, {\it J. Math.\ Phys.} {\bf 19}, 1156--1162
\item  Miura R M, Gardner C S and Kruskal M D 1968, {\it J. Math.\ Phys.} {\bf
9}, 1204--1209
\item  Oevel G and Fuchssteiner B 1992, {\it Physica A} {\bf 181}, 364--384
\item  Olver P J 1980, {\it Proc.\ Camb.\ Phil.\ Soc.} {\bf 88}, 71--88
\item  Olver P J 1986, {\it Applications of Lie Groups to Differential
Equations} (Springer, N Y)
\item  Rogister A 1971, {\it Phys.\ Fluids} {\bf 14}, 2733--2739
\item  Spangler S R 1992, in: {\it Solar Wind Seven} (edited by E Marsch and
R Schwenn, Pergamon Press, Oxford) 539--544
\item  Spangler S R and Plapp B B 1992, {\it Phys.\ Fluids B} {\bf 4},
3356--3370
\item  Verheest F and Hereman W 1994, {\it Physica Scripta}, in press

\end{description}

\end{document}